\documentclass[fleqn,12pt,twoside]{article}
\usepackage{espcrc1}
\usepackage{graphicx}
\usepackage[figuresright]{rotating}

\newcommand{\AmS}{{\protect\the\textfont2
  A\kern-.1667em\lower.5ex\hbox{M}\kern-.125emS}}

\title{The $^4$He($\gamma,n$)$^3$He Reaction with Full Final State Interaction}

\author{S. Quaglioni\address[MCSD]{Dipartimento di Fisica, 
	Universit\`a di Trento and INFN (Gruppo Collegato di Trento),\\ 
        via Sommarive 14, I-38050 Povo (Trento), Italy}, %
	N. Barnea\address{The Racah Institute of Physics, The Hebrew
	Universty,\\ 91904, Jerusalem, Israel },
	V. D. Efros\address{Russian Research Centre ``Kurchatov Institute'',\\
	Kurchatov Square 1, 123182 Moskow, Russia},
        W. Leidemann\addressmark[MCSD] 
        and
        G. Orlandini\addressmark[MCSD]%\thanks{To reuse an addressmark
	}
       
\begin{document}

% typeset front matter
\maketitle
The total cross sections of the processes $^{4}$He$\left(\gamma,p\right)^{3}$H and $^{4}$He$\left(\gamma,n\right)^{3}$He are calculated. For these exclusive reactions we investigate the 
question of the giant dipole peak height, but we also consider higher energies. The calculation includes full Final State Interaction (FSI) via the Lorentz Integral 
Transform (LIT) approach \cite{ELO94} and employs the semi-realistic MTI-III potential \cite{MT}. 
The LIT method has already been successfully applied to the calculation of the exclusive $d(e,e^{\prime}p)n$ reaction \cite{LL00}. We would like to emphasize that FSI is taken rigorously into account also in the region beyond the three-body break-up threshold.
From the total photoabsorption cross section for the same potential calculated in \cite{ELO97,BELO01} we are also able to determine the sum of three- and four-body break-up cross sections.
Here (as well as in \cite{BELO01,ELO97}) only the transitions induced by the unretarded dipole operator $D$ are taken into account.
The total exclusive cross section of the photodisintegration of $^4$He into the two-fragments ($\mathit{N,3}$), where {\it{N}} represents the scattered proton (neutron) and {\it 3} the $^3$H ($^3$He) nucleus, is given by
\begin{equation}
\sigma_{(\gamma,\mathit{N})}=\frac{e^2}{\hbar c}\omega_{\gamma} k\mu\int{\left|\left\langle\Psi_{\mathit{N;3}}^-(E_f)\left|D\right|\Psi_{\alpha}\right\rangle\right|^2 d\Omega_k~,~~~ E_f=\omega_{\gamma}+E_{\alpha}}~.
\end{equation}
With $\mu$ and $k$ we indicate the reduced mass and the relative momentum of the two fragments, respectively; $\omega_{\gamma}$ is the incident photon energy, $E_{\alpha}$ and $\Psi_{\alpha}$ are energy and wave function of the $\alpha$ particle bound state, whereas
\begin{equation}
\left|\Psi^{-}_{\mathit{N;3}}(E_f)\right\rangle=\widehat{\mathcal A}\left|\phi_1(E_f)\right\rangle+\frac{1}{E_f-i\varepsilon-H}\widehat{\mathcal A}V_1\left|\phi_1(E_f)\right\rangle
\end{equation}
is the formal scattering solution of the ({\it{N,3}}) channel at energy $E_f$: $\phi_1$ is the unperturbed wave function describing the relative motion of nucleon 1 with respect to nucleons 2, 3 and 4 bound to form nucleus {\it 3}, $V_1$ is the corresponding {\it{N-3}} interaction part of the full nuclear Hamiltonian $H$ and $\widehat{\mathcal A}$ is the antisymmetrization operator. The Coulomb interaction is taken into account both by using Coulomb wave functions for $\phi_1$ in the  ($p,^3$H) channel (instead of spherical Bessel functions for ($n,^3$He)), and in the calculation of the bound states.
In the LIT method the various transition matrix elements are  calculated using the relation \cite{LL00,Ef85}: 
\begin{equation}
\langle \Psi^-_{\mathit{N;3}}\left(E_{f}\right)\left| D \right|\Psi_{\alpha} \rangle = \left\langle \phi_1\left(E_{f}\right)\left|{\widehat{\mathcal A}} D \right|\Psi_{\alpha} \right\rangle + \left\langle \phi_1\left(E_{f}\right)\left|V_1{\widehat{\mathcal A}}\frac{1}{E_f+i\varepsilon-H} D \right|\Psi_{\alpha}\right\rangle~.
\end{equation}
The first matrix element, $\left\langle \phi_1\left(E_{f}\right)\left| {\widehat{\mathcal A}}D \right|\Psi_{\alpha} \right\rangle$, 
can be calculated without greater problems. The second matrix element, which represents the difficult 
problem, is calculated as follows. One defines the function
\begin{equation}
F(E) = \int{df\langle\phi_1|V_1{\widehat{\mathcal A}}|\Psi_f(E^{\prime})\rangle\langle\Psi_f(E^{\prime})|D|\Psi_{\alpha}\rangle \delta(E-E^{\prime})} \,,
\end{equation}
where $\Psi_f$ is a complete set of Hamiltonian eigenstates, then:
\begin{equation}
\left\langle \phi_1\left(E_{f}\right)\left|V_1{\widehat{\mathcal A}}\frac{1}{E_f+i\varepsilon-H} D \right|\Psi_{\alpha}\right\rangle=-i\pi F(E_f)+{\mathcal P}\int_{E_{th}}^{\infty}{\frac{F(E)}{E_f-E}~dE}~.
\end{equation} 
The function $F$ is computed indirectly  via its LIT. One can show \cite{Mar02}
 that the LIT of $F$ can be written in terms of a Lanczos orthonormal basis $\{\left|\varphi_i\right\rangle,i=0, ..., n\}$,
 \begin{equation}
{\mathcal L}[F]=-\frac{\sqrt{\left\langle\Psi_{\alpha}\left|DD\right|\Psi_{\alpha}\right\rangle}}{\sigma_{I}}\sum_{i=0}^n{\left\langle\phi_1\left|V_1{\widehat\mathcal A}\right|\varphi_i\right\rangle}\Im\left\{\left\langle\varphi_i\left|\frac{1}{\sigma-H}\right|\varphi_o\right\rangle\right\},~~~\sigma=\sigma_{R}+i\sigma_I,
\end{equation}
and the matrix elements $\langle\varphi_i|(\sigma-H)^{-1}|\varphi_o\rangle$ can be written as continued fractions of the Lanczos coefficients.
\begin{figure}[!t]
\vspace{-5mm}
$\!\!\!\!\!$\raggedright\includegraphics[scale=0.56]{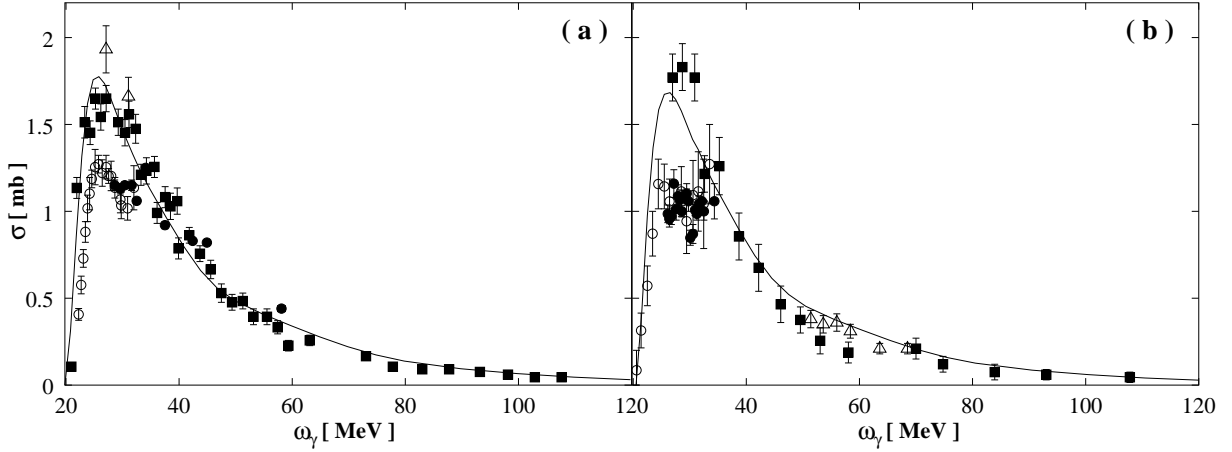}
\vspace{-15mm}
\caption{Figure (a) shows the result for the $^4\rm{He}(\gamma,\rm{p})^3\rm{H}$ cross section (solid curve) in comparison with experimental data of \cite{Ark74} (squares), \cite{Cal83} (triangles), \cite{Ber88} (full circles)  and \cite{Fel90} (open circles); Figure (b) shows the result for the $^4\rm{He}(\gamma,\rm{n})^3\rm{He}$ cross section (solid curve) in comparison with experimental data of \cite{Ark74} (squares), \cite{Ber80} (open circles), \cite{War81} (full circles) and \cite{Sim98} (triangles).}
\label{fig1}
\vspace{-2mm}
\end{figure}

\begin{figure}[!t]
\vspace{-6mm}
\centering\includegraphics[scale=0.56]{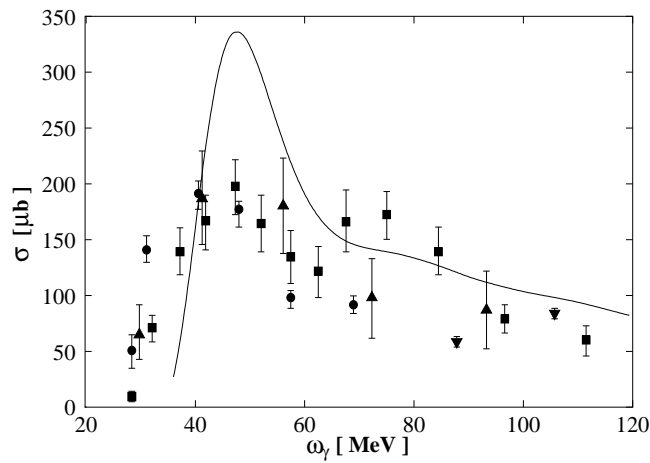}
\vspace{-10mm}
\caption{Difference between the total cross section of \cite{BELO01} and the present ($\gamma,p$) and ($\gamma,n$) results in comparison with $^4{\rm He}(\gamma,pn)d$ experimental data of \cite{Gor58} (upward triangles), \cite{Ark70} (squares), \cite{Bal77} (circles), \cite{Dor93} (downward triangles).}
\label{fig2}
\vspace{-2mm}
\end{figure}

The obtained cross sections for ($\gamma,p$) and ($\gamma,n$) channels are presented in Figure \ref{fig1} in comparison with experimental data. In both cases the theoretical results show a pronounced giant dipole peak tending towards the data of \cite{Ark74} and \cite{Cal83} rather than showing the flat behavior of the data of \cite{Ber88,Fel90,Ber80,War81}. Beyond the peak one finds a rather good agreement with the data of \cite{Ark74} and \cite{Sim98}. In Figure \ref{fig2} we show the difference between the total cross section of \cite{BELO01} and the ($\gamma,p$) and ($\gamma,n$) ones compared to the three-body break-up data. The comparison between theory and experiments turns out to be quite reasonable, even though it is evident that more work is needed on the experimental side as well as on the theoretical side using more realistic interactions.


\begin{thebibliography}{00}

\bibitem{ELO94} V.D. Efros, W. Leidemann, and G. Orlandini, Phys. Lett.
B338 (1994) 130.
\bibitem{MT} R.A. Malfliet and J. Tjon, Nucl. Phys. 127 (1969) 161.
\bibitem{LL00} A. La Piana and W. Leidemann, Nucl. Phys. A677 (2000) 423.
\bibitem{ELO97} V.D. Efros, W. Leidemann, and G. Orlandini, Phys. Rev. Lett. 78 (1997) 4015.
\bibitem{BELO01} N. Barnea, V.D. Efros, W. Leidemann, and G. Orlandini,
Phys. Rev. C63 (2001) 057002.
\bibitem{Ef85} V.D. Efros, Sov. J. Nucl. Phys. 41 (1985) 949; Phys.
Atom. Nucl. 62 (1999) 1833.
\bibitem{Mar02} M.A. Marchisio, N. Barnea, W. Leidemann and G. Orlandini, nucl-th/0202009, Few Body Sys. in print.
\bibitem{Ark74} Yu.M. Arkatov et al., Sov. J. Nucl. Phys. 19(6), 598 (1974).
\bibitem{Cal83} J.R. Calarco et al., Phys. Rev. C28, 483 (1983).
\bibitem{Ber88} R. Bernabei et al., Phys. Rev. C38, 1990 (1988).
\bibitem{Fel90} G. Feldman et al., Phys. Rev. C42, 1167 (1990).
\bibitem{Ber80} B.L. Berman, D.D. Faul, P. Meyer, and D.L. Olson,
Phys. Rev. C22 (1980) 2273.
\bibitem{War81} L. Ward et al. Phys. Rev. C24, 317 (1981).
\bibitem{Sim98} D.A. Sims et al., Phys. Lett. B442, 43 (1998).
\bibitem{Gor58} A.N. Gorbunov et al., Sov. Phys. JEPT 34, 600 (1958); Sov. J. Nucl. Phys. 10, 268 (1969).
\bibitem{Ark70} Yu.M. Arkatov et al., Sov. J. Nucl. Phys. 10, 639 (1970).
\bibitem{Bal77} F. Balestra et al., Nuovo Cim. 49A, 575 (1979); Nuovo Cim. 38A, 145 (1977).
\bibitem{Dor93} S.M. Doran et al. Nucl. Phys. A559, 347 (1993).
\end{thebibliography}
\end{document}